\documentclass[apjl]{emulateapj}
\usepackage{apjfonts}

\usepackage{natbib}
\usepackage{graphicx}
\usepackage{epsfig}
\usepackage{amsmath}
\newcommand{\nW}{nW m$^{-2}$ sr$^{-1}$}

\shorttitle{CONTRIBUTION OF LENSED SCUBA GALAXIES TO THE CIB}
\shortauthors{ZEMCOV ET AL.}

\begin{document}

\slugcomment{Submitted to ApJ 2010 March 17; accepted 2010 July 28;
  published 2010 August 27}
\title{Contribution of lensed SCUBA galaxies to the cosmic infrared background}

\author{Michael Zemcov\altaffilmark{1,2},
Andrew Blain\altaffilmark{1},
Mark Halpern\altaffilmark{3},
Louis Levenson\altaffilmark{1}
}
\altaffiltext{1}{Division of Physics, Mathematics \& Astronomy,
  California Institute of Technology, Pasadena, CA 91125, USA.} 
\altaffiltext{2}{Jet Propulsion Laboratory, 4800 Oak Grove Drive,
  Pasadena, CA 91109, USA} 
\altaffiltext{3}{Department of Physics \& Astronomy, University of
  British Columbia, Vancouver, BC V6T 1Z1, Canada} 

\begin{abstract}
  The surface density of submillimeter (sub-mm) galaxies as a function of
  flux, usually termed the source number counts, constrains models of
  the evolution of the density and luminosity of starburst galaxies.
  At the faint end of the distribution, direct detection and counting
  of galaxies are not possible.  However, gravitational lensing by
  clusters of galaxies allows detection of sources which would
  otherwise be too dim to study.  We have used the largest catalog of
  sub-mm-selected sources along the line of sight to galaxy clusters
  to estimate the faint end of the $850 \, \mu$m number counts;
  integrating to $S = 0.10 \,$mJy the equivalent flux density at $850
  \, \mu$m is $\nu I_{\nu} = 0.24 \pm 0.03 \,$\nW.  This provides a
  lower limit to the extragalactic far-infrared background and is
  consistent with direct estimates of the full intensity from the
  FIRAS.  The results presented here can help to guide strategies for
  upcoming surveys carried out with single dish sub-mm instruments.
\end{abstract}



\section{Introduction}
\label{S:intro}

\setcounter{footnote}{0}

Measurements of the sources which comprise the cosmic infrared
background (CIB) radiation provide constraints on the most vigorous
epoch of star formation in the universe.  Deep submillimeter (sub-mm)
surveys using the Sub-mm Common-User Bolometer Array (SCUBA;
\citealt{Holland1999}) have detected a population of high redshift,
ultraluminous sub-mm galaxies with star formation rates approaching
$1000 \, M_{\sun}$ per year (\citealt{Smail1997};
\citealt{Hughes1998}; \citealt{Barger1998}; \citealt{Blain1999};
\citealt{Barger1999}; \citealt{Eales2000}; \citealt{Cowie2002};
\citealt{Scott2002}; \citealt{Chapman2002}; \citealt{Smail2002};
\citealt{Webb2003}; \citealt{Borys2003}; \citealt{Coppin2006};
\citealt{Knudsen2006}; \citealt{Knudsen2008}).  Such surveys have been
very successful in characterizing the number counts of 850$\mu$m
sources with fluxes $\gtrsim 2 \,$mJy, and have resolved as much as 50
\% of the extragalactic background light at these wavelengths
(e.g.~\citealt{Coppin2006}).  However, due to SCUBA's confusion noise
of $\sim 0.5 \,$mJy at $1\sigma$ and a source density which rises
steeply with decreasing flux, obtaining constraints on the number
counts of sources at the low fluxes corresponding to the bulk of the
CIB has been difficult.  \citet{Blain1997} advocated using
gravitational lensing by galaxy clusters to amplify the brightness of
these sources above the James Clerk Maxwell Telescope (JCMT) confusion
limit.  This method was first used in pioneering work with SCUBA data
by \citet{Smail1997} and subsequently by others (\citealt{Smail2002};
\citealt{Cowie2002}; \citealt{Knudsen2006}; \citealt{Knudsen2008}), to
measure the background contribution from faint point sources at high
redshifts, as well as the shape of the point source counts down to
$\sim 0.1 \,$mJy.  Unfortunately, these studies suffer from large
Poisson noise due to a small number of sources in their sample as well
as challenging analysis issues; a larger data set is desirable.

Models of the evolution of star-forming galaxies are fit to counts of
galaxies at wavelengths ranging from $24 \, \mu$m to $2 \,$mm (for
example \citealt{Lagache2003}; \citealt{Negrello2007};
\citealt{Valiante2009}).  The counts observed at $850 \, \mu$m play a
particularly important role because they span a broad range in
redshift, across which episodic starburst activity is believed to have
varied widely.  It is therefore particularly important that the counts
be extended to faint enough fluxes to capture the bulk of the CIB, as
has now been done at shorter wavelengths by BLAST
(\citealt{Patanchon2009}; \citealt{Marsden2009}).

The slope of the $850 \, \mu$m source surface density must break to a
flatter slope than is measured for bright sources
(e.g.~\citealt{Coppin2006}) in order that the total flux remain
consistent with absolute measurements of the CIB (\citealt{Puget1996};
\citealt{Fixsen1998}).  The precise shape of this break will constrain
models of the evolution of number density and luminosity of fainter
infrared-luminous dusty galaxies and will also provide clues about the
relation of this population to more ordinary, luminous sub-mm
galaxies.

In this paper, we utilize the SCUBA Galaxy Cluster Survey from
\citet{Zemcov2007}, the largest $850 \, \mu$m galaxy cluster data set
to date, and a model of the lensing in each cluster to form a catalog
of sources on the line of sight through the clusters.  These sources
allow the most reliable constraints available on the faint source
contribution to the CIB at $850 \, \mu$m.  Knowledge of the amplitude
and shape of the source density spectrum will help set the stage for
new sub-mm experiments like SCUBA-2, LABOCA, and {\sc herschel} which
are capable of mapping hundreds of lensing clusters over unprecedented
sky areas.


\section{Data sample and analysis pipeline}
\label{S:analysis}

\subsection{Survey sources}
\label{sS:survey}

More than 40 clusters were mapped over SCUBA's operational lifetime
with integration times ranging from $1 \,$ks to $> 100 \,$ks; all those
archival SCUBA data taken as part of programs to map galaxy clusters
comprise the data set presented in \citet{Zemcov2007}.  The initial
catalog used in this analysis consists of the $850 \, \mu$m sources
listed in Table 5 of \citet{Zemcov2007}; that paper gives a complete
discussion of the data sample, low level analysis, map making and
source extraction techniques.

It is thought that most blank field extragalactic SCUBA sources are at
high redshift \citep{Lagache2005} and that their flux is thermal dust
emission heated by an evolving combination of star formation and
active galactic nuclei.  We expect this holds for the sources in this
work with two classes of exception: those sources known to be situated
inside the cluster under investigation (e.g.~Abell 780-1) and those
\emph{apparent} sources which may instead be images of the
Sunyaev-Zel'dovich (SZ) effect (e.g.~Abell 1689-2), which is positive
at $850 \, \mu$m.  The former are radio bright so cross
identifications are easily found in the literature; such sources are
readily separated from the high-redshift sample.  The latter class of
source is necessarily near the center of the cluster where background
sources are \emph{demagnified}.  We expect no sources behind the
cluster center to be detected and treat all central sources as cluster
members.  These two classes of sources are discussed in detail in
\citet{Zemcov2007}. In this work, we assume that both types are
sources \emph{in} the cluster (that is, not gravitationally lensed
\emph{by} it). The flux from such sources can simply be summed without
magnification to yield their contribution to the total flux density.

The full \citet{Zemcov2007} sample contains some maps which are
extremely noisy.  We excise from the full sample all maps where the
pixel rms in a SCUBA beam is $> 10 \,$mJy and those maps where the
data are obviously corrupted by instrumental problems.  We are left
with a subsample of 28 clusters with which to measure the contribution
to the CIB; these are listed in Table \ref{tab:vdz}.

\begin{table}
\centering
\caption{Cluster Fields, Redshifts and Velocity Pispersion Parameters used in this Analysis.}
\begin{tabular}{lccc}
\hline
Cluster & $z$ & $v_{\mathrm{z}}$ & Reference \\ \hline
Abell 209 & 0.209 & $1394^{+88}_{-99}$ & \citet{Mercurio2003} \\
Abell 370 & 0.375 & $1364^{+50}_{-50}$ & Struble et al.~(1991) \\
Abell 383 & 0.187 & $900^{+10}_{-10}$ & \citet{Smith2005} \\
Abell 478 & 0.088 & $904^{+261}_{-140}$ & \citet{Wu1999} \\
Abell 496 & 0.033 & $687^{+89}_{-76}$ & \citet{Fadda1996} \\
Abell 520 & 0.199 & $1250^{+189}_{-189}$ & \citet{Proust2000} \\
Abell 586 & 0.171 & $1161^{+196}_{-196}$ & \citet{Cypriano2005} \\
Abell 773 & 0.217 & $750^{+60}_{-70}$ & \citet{Smith2005} \\
Abell 851 & 0.407 & $1081^{+31}_{-31}$ & \citet{Goto2003} \\
Abell 963 & 0.206 & $980^{+15}_{-15}$ & \citet{Smith2005} \\
Abell 1689 & 0.183 & $1290^{+100}_{-100}$ & \citet{Lokas2006} \\
Abell 1835 & 0.253 & $1210^{+80}_{-100}$ & \citet{Smith2005} \\
Abell 2204 & 0.152 & $1029^{+72}_{-59}$ & \citet{Pimbblet2006} \\
Abell 2218 & 0.176 & $1070^{+5}_{-5}$ & \citet{Smith2005} \\
Abell 2219 & 0.226 & $902^{+10}_{-10}$ & \citet{Smith2005} \\
Abell 2390 & 0.228 & $1100^{+80}_{-80}$ & Natarajan et al.~(2004) \\
Abell 2597 & 0.085 & $776^{+101}_{-101}$ & \citet{Cypriano2004} \\
Cl$\, 0016{+}16$ & 0.541 & $984^{+130}_{-95}$ & \citet{Poggianti2006} \\
Cl J$0023{+}0423$A & 0.827 & $415^{+102}_{-63}$ & \citet{Lubin1998} \\
Cl$\, 0024.0{+}1652$ & 0.390 & $1000^{+70}_{-70}$ & Natarajan et al.~(2004) \\
ClG J$0848{+}4453$ & 1.27 & $640^{+90}_{-90}$ & \citet{Rosati1999} \\
MS$\, 1455.0{+}2232$ & 0.258 & $1032^{+130}_{-95}$ & \citet{Borgani1999} \\
Cl J$1604{+}4304$ & 0.895 & $982^{+100}_{-100}$ & \citet{Crawford2006} \\
Cl$\, 2244{-}0221$ & 0.330 & $600^{+80}_{-80}$ & Natarajan et al.~(2004) \\
MS$\, 0440.5{+}0204$ & 0.190 & $715^{+113}_{-68}$ & \citet{Borgani1999} \\
MS$\, 0451.6{-}0305$ & 0.550 & $1330^{+111}_{-94}$ & \citet{Borgani1999} \\
MS$\, 1054.4{-}0321$ & 0.823 & $1170^{+150}_{-150}$ & \citet{Tran1999} \\
RX J$1347.5{-}1145$ & 0.451 & $1500^{+160}_{-160}$ & Fischer et al.~(1997) \\
\hline
\end{tabular}
\label{tab:vdz}
\end{table}

A direct estimate of the total background arising from lensed sources
can be made without the intermediate step of constructing demagnified
number counts by exploiting the fact that gravitational lenses are
like any other optical system in that the \'etendue of the system, $A
\Omega$, is conserved; this means that, integrating over all $\Omega$,
the total flux is equal before and after passing through the lens. The
flux density can therefore be computed by integrating all of the flux
in the sources in this survey and dividing by the area observed; this
procedure yields the CIB lower limit $\nu I_{\nu} \geq 0.19 \,$\nW.
Unfortunately, this method is limited because the $\Omega$ and depth
observed by SCUBA in these 28 clusters is significantly smaller and
shallower than that required to contain all the flux from the reported
{\sc firas} background.  We therefore expect this constraint to be
smaller than the result from a more comprehensive analysis.
Furthermore, it is difficult to assign an uncertainty to this approach
as the survey completeness is not a simple function of the measurement
error in the image plane and certainly depends on the true shape of
the source counts.  We therefore employ a gravitational lens model to
make progress.

\subsection{Gravitational Lensing Demagnification}
\label{sS:gravlens}

We model the lensing of each cluster using the mass distribution
associated with the Navarro-Frenk-White (NFW) model
\citep{Navarro1996} since that has been shown to be a very good
description of systems whose mass is dominated by dissipationless dark
matter.  The NFW model is discussed in works like \citet{Wright2000}
and \citet{Li2002} which provide all of the materials necessary to
calculate the deflections and magnifications of idealized lensing
clusters as a function of the concentration parameter $c(z)$ and
velocity dispersion $\sigma_{v}$.

The background source counts are determined from the catalog as
follows.  First, the catalog fluxes and the survey completeness are
corrected for the effects of gravitational lensing using the lens
model.  The de-lensed flux counts are then completeness corrected and
summed in bins to determine the measured cumulative source counts
$N_{\mathrm{meas}}(>S)$.  Simulations are performed using different
input source count models to compute the effects of flux boosting
(Eddington bias; see, e.g., \citealt{Eales2000}) on the true,
underlying background source counts $N_{\mathrm{true}}(>S)$;
comparison of the simulated, flux-boosted counts
$N_{\mathrm{sim}}(>S)$ with $N_{\mathrm{meas}}(>S)$ yields the
corresponding $N_{\mathrm{true}}(>S)$ and these lead to a measurement
of the total CIB intensity.

Given an NFW model for each cluster in the survey, the magnification
at each background pixel in the image plane map can be computed.  The
magnifications derived from the lensing model are applied to the
measured source catalog to determine the source plane flux of each
of the sources.  The model is different for each cluster and the
demagnification factor for each source depends on its position.
Similarly, to compute the completeness of the survey, the source plane
noise maps are divided by the same set of model magnifications to
create a background error map in which the value at each pixel is the
effective measurement error at the source plane.  These maps are
multiplied by the source detection threshold value from
\citet{Zemcov2007}, and the resulting pixel values are summed to give
the completeness as a function of source flux, $C(>S)$: the $50$\%
completeness of this survey, averaged over all fields, is $\sim 5
\,$mJy in the source plane.  The distribution of source plane fluxes
is then corrected using $C(>S)$ and summed into bins to yield
$N_{\mathrm{meas}}(>S)$.

The uncertainties in the derived $N_{\mathrm{meas}}(>S)$ depend on the
measurement errors of the catalog, the uncertainty associated with the
lensing models, and cosmic variance in the source sample.  The most
concise method of accounting for all of these sources of uncertainty
is to use a Monte Carlo approach.  The input parameters are varied
over their nominal uncertainty range; these lead to a set of
$N_{\mathrm{meas}}(>S)$ which span the allowed range of output
values given the input uncertainties.

Determining the uncertainty due to cosmic variance is straight
forward: in each simulation, a random realization of sources
consistent with the counts is assigned uncorrelated random
positions before the lensing model is applied.  Over many such
simulations, variations in $N_{\mathrm{meas}}(>S)$ due to cosmic
variance should be measured to high accuracy.  In each of the
uncertainty simulations discussed below, cosmic variance is included.

For each source, the \citet{Zemcov2007} catalog assigns a measurement
error whose effect on $N_{\mathrm{meas}}(>S)$ is determined as
follows.  Each source's nominal flux is varied by an amount drawn from
a Gaussian probability density function (PDF) with standard deviation
equal to the measurement error of the source.  This modified flux is
then demagnified according to the model to determine its equivalent
source plane value.

Estimating the full uncertainty requires that the lens model
parameters be varied on a per cluster basis. One of the least
constrained parameters is the source redshift.  The difficulty
associated with determining unique counterparts of sub-mm sources at
other wavelengths, whether in galaxy clusters or not, has rendered the
precise redshifts for most of the catalog sources uncertain or
unmeasured.  As lensing magnification is not a strong function of
source redshift beyond $z \sim 1$ \citep{Blain1997}, the approach
utilized in this analysis is to draw from a source redshift PDF based
on the $n(z)$ results presented in \citet{Aretxaga2007}.  In each
calculation, the source plane for each cluster is assigned a redshift
drawn at random from this distribution.  Over many calculations, the
full deflector-source redshift $z_{\mathrm{ds}}$ parameter space is
probed.  The set of $z_{\mathrm{ds}}$ could be varied on a per source
basis; here, for computational efficiency, we choose to vary it per
field.  Over many realizations this is an equivalent procedure.  As
these are well-studied clusters with precisely known redshifts, the
cluster redshifts are never varied.  A second parameter which is
varied is the cluster's velocity dispersion, $\sigma_{v}$.
These are typically available in the literature and are used here as
the mean and standard deviations of Gaussian distributions from which
$\sigma_{v}$ is randomly drawn; the scatter in the results of
such realizations is a measurement on the effect of uncertainty in the
lens model on $N_{\mathrm{meas}}(>S)$.

Though each of these parameters could be varied alone or in any
combination, all are varied simultaneously as we determine the
uncertainty in the measurement of $N_{\mathrm{meas}}(>S)$.  This
accurately reflects the lack of covariance between the different model
parameters.  In the results presented here, $10^{4}$ simulations were
performed and propagated through to $N_{\mathrm{meas}}(>S)$.  In each
flux bin, the uncertainties in $N_{\mathrm{meas}}(>S)$ are determined
by calculating the set of $N(>S)$ which encompass $68$\% of the
simulation results about the most likely value and taking the extrema
of the set.

\begin{figure}
\centering
\epsfig{file=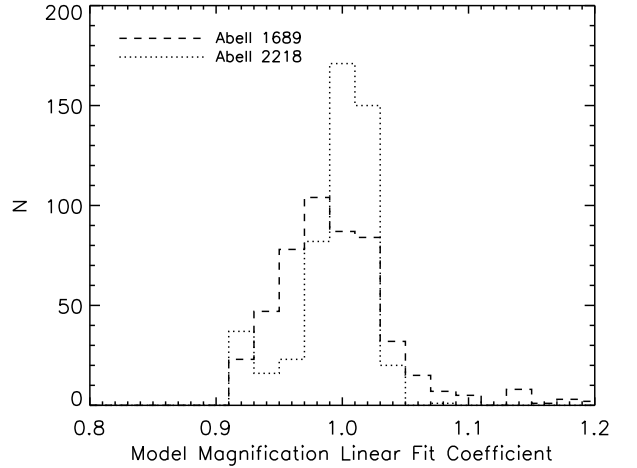,width=0.49\textwidth}
\caption{Histogram of linear fit coefficients arising from a
  comparison of the magnified fluxes derived from the NFW model used
  in this work and the {\sc lenstool} models which are publicly
  available for Abell 1689 and Abell 2218.  These distributions,
  derived from $500$ simulations, are scattered around 1.0 with a
  standard deviation of $\pm 0.07$ for Abell 1689 and $\pm 0.04$ for
  Abell 2218.  This shows that the NFW model can be used as an
  accurate and unbiased statistical description of a set of
  magnifications on many backgrounds in a given cluster.}
\label{fig:mu}
\end{figure}

Many previous source count results have been derived from detailed
lensing models of clusters.  However, these models are observationally
expensive and time consuming to construct and, unlike simple
observables like velocity dispersions, are typically not available in
the public realm.  Two clusters in our sample which do have publicly
available lensing models written using the {\sc lenstool} code
\citep{Jullo2007} are Abell 1689 \citep{Limousin2007} and Abell 2218
\citep{Eliasdottir2007}.  To compare the statistical properties of the
NFW model to these more detailed models, we perform simulations where
backgrounds, populated using random realizations of the source counts
in the way described above, are lensed with both the {\sc lenstool}
and NFW models.  The set of magnified source fluxes (of which only a
very few are magnified by a factor greater than a few) can be fit to
one another to yield a linear scaling between the resulting flux
populations.  Over many realizations, if the NFW model is capturing
the statistical behavior of the lensing magnification as modeled by
the {\sc lenstool} model, the scaling between them should be equal to
1.  As the NFW model can lead to very large magnifications, we set a
threshold maximum magnification of a factor of 25; sources whose
magnification factor is larger than this are set to this limit.
Figure \ref{fig:mu} shows the results of $500$ such simulations for
both Abell 1689 and Abell 2218 models; the mean relation is indeed
consistent with unity and the derived fit slopes have standard
deviations of about 5\% in both clusters.  Uncertainties in the input
velocity dispersions move the mean of the distributions within this
standard deviation; presumably, a biased velocity dispersion
measurement would also bias the lensing model though the bias would
have to be $> 100 \,$km s$^{-1}$ to have much effect. This is evidence
that, when used in Monte Carlo simulations, the NFW model accurately
captures the bulk magnification properties of the more expensive
models.

\subsection{Flux Bias Correction}
\label{sS:boosting}

Though computation of $N_{\mathrm{meas}}(>S)$ and its errors does not
require any a priori assumptions about the shape of
$N_{\mathrm{true}}(>S)$, correcting $N_{\mathrm{meas}}(>S)$ for the
effects of flux boosting to determine $N_{\mathrm{true}}(>S)$ does;
the effects of flux boosting are estimated as follows.  As a model
parameterizing $N(>S)$ we use the broken power law
\begin{equation}
\label{eq:NgtS}
\begin{split}
\frac{dN}{dS} & = N_{0} \left( \frac{S}{S_{0}} \right)^{-\beta}
  \: \mathrm{for} \: S > S_{0}, \\
\frac{dN}{dS} & = N_{0} \left( \frac{S}{S_{0}} \right)^{-\alpha}
\: \mathrm{for} \: S \leq S_{0}.
\end{split}
\end{equation}
$N(>S)$ is the cumulative integral of Equation (\ref{eq:NgtS}) from
infinite flux to flux $S$.  This broken power-law functional form for
$N(>S)$ has a long pedigree in SCUBA blank field studies and, although
not physically motivated, fits the current data above the confusion
limit very well (see, e.g., \citet{Coppin2006}).  Since the
total source plane area of this survey is not large, there are few
intrinsically bright sources in it.  To construct a model for source
de-boosting, we therefore fix the model parameters which are best
measured using sources with $S > 10 \,$mJy, namely $\beta$ and
$S_{0}$, to the values found by \citet{Coppin2006}, $5.1$ and $9
\,$mJy, respectively, and only ever vary $\alpha$ and $N_{0}$.

At a fixed $\alpha$ and $N_{0}$ a set of sources whose flux
distribution is consistent with Equation (\ref{eq:NgtS}) are assigned
Gaussian distributed random positions within a square grid.  These
maps are then lensed using the set of NFW cluster mass profiles.  The
calculation of flux boosting in SCUBA data is complicated by the fact
that SCUBA is a double-differencing instrument \citep{Coppin2006}.
To account for this, each simulated source's flux is modified
according to:
\begin{equation}
\label{eq:boost}
S'  =  S(\alpha_{\mathrm{c}},\delta_{\mathrm{c}}) -
\frac{1}{2} [S(\alpha_{\mathrm{l}},\delta_{\mathrm{l}}) +
S(\alpha_{\mathrm{r}},\delta_{\mathrm{r}})]
+ \sigma
\end{equation}
Here $S$ is the flux in the simulated image plane map at position
$(\alpha,\delta)$ where `c' denotes the on source chop position and
`l' and `r' denote the two chopped positions, $\sigma$ is a
realization of noise drawn from a Gaussian PDF with standard deviation
equal to that measured in the SCUBA map, and $S'$ is the biased flux
of the source.  To reduce the complexity of these simulations, the
SCUBA chop throw is set to $60''$, which is the median of the
values used in the \citet{Zemcov2007} sample.  The flux of each source
in the simulated catalog is varied according to this algorithm, and the
simulated source counts $N_{\mathrm{sim}}(>S)$ are formed by summing
into bins in the usual way.  As the simulated sources are boosted in
the same way as the real sources and the simulated and measured
catalogs are identically binned, $N_{\mathrm{sim}}(>S)$ can be
directly compared with the boosted measured source counts
$N_{\mathrm{meas}}(>S)$.

Simulated, flux-boosted counts are calculated for a range of the
parameters $\alpha$ and $N_{0}$; the outputs of such simulations,
$N_{\mathrm{sim}}^{\alpha,N_{0}}(>S)$, are then directly compared to
$N_{\mathrm{meas}}(>S)$.  For each pair of $(\alpha,N_{0})$, 100
simulations are performed to measure the mean and the scatter in
$N_{\mathrm{sim}}^{\alpha,N_{0}}(>S)$ arising from the variation in
different noise realizations of Equation (\ref{eq:boost}); the variance
in the $N_{\mathrm{sim}}^{\alpha,N_{0}}(>S)$ is typically 2 orders
of magnitude smaller than the uncertainty in $N_{\mathrm{meas}}(>S)$
and does not significantly contribute to the overall uncertainty.  For
each mean $N_{\mathrm{sim}}^{\alpha,N_{0}}(>S)$ value,
\begin{equation}
\label{eq:chisq}
\chi^{2} = \sum \left[ N_{\mathrm{meas}}(>S) -
N_{\mathrm{sim}}^{\alpha,N{0}}(>S) \right]^{2} / \sigma_{\mathrm{meas}}^{2}, 
\end{equation}
where the sum is over flux bins.  This is converted to a
likelihood function which allows us to assess the agreement between
the different model realizations and $N_{\mathrm{meas}}(>S)$.  

To correct $N_{\mathrm{meas}}(>S)$ for flux boosting, the model with
$(\alpha, N_{0})$ yielding the most probable fit to the data is
determined, and for each flux bin the ratio of the boosted to input
$N(>S)$ is calculated and used to correct the $N_{\mathrm{meas}}(>S)$
to determine $N_{\mathrm{true}}(>S)$.  


\section{Results and discussion}
\label{S:results}

\begin{figure}
\centering
\epsfig{file=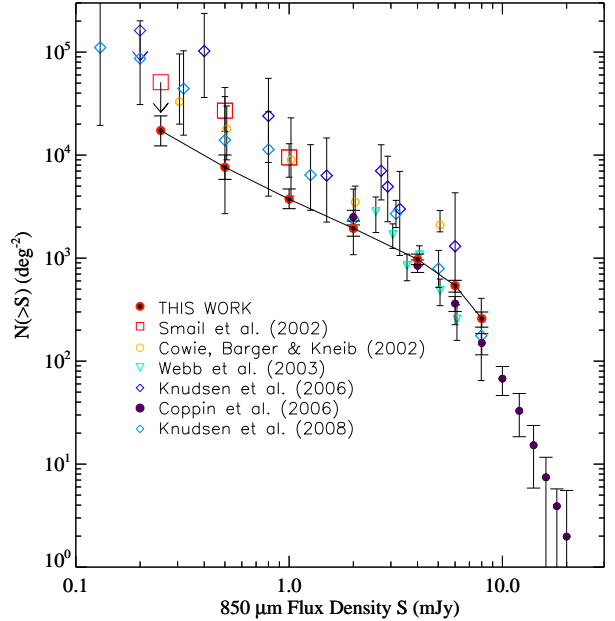,width=0.49\textwidth}
\caption{Cumulative source counts $N(>S)$ derived from the
  \citet{Zemcov2007} catalog using the analysis described in the
  text (red points).  The error bars account for the uncertainties
  associated with source flux measurement errors, gravitational
  lensing model uncertainties, flux boosting and cosmic variance.  The
  other points are from the following sources: \citet{Smail2002},
  \citet{Cowie2002}, \citet{Webb2003}, \citet{Knudsen2006}, 
  \citet{Coppin2006}, and \citet{Knudsen2008}.}
\label{fig:NgtS}
\end{figure}

The isotropic cumulative source counts $N_{\mathrm{true}}(>S)$ for
fluxes below $10 \,$mJy inferred by our catalog are shown as filled
dots connected by a solid line in Figure \ref{fig:NgtS} along with
prior measurements of the density of $850 \, \mu$m selected sources.
Table \ref{tab:meas} lists the measurements plotted in Figure
\ref{fig:NgtS} for reference.  Our results are consistent with prior
data taken above SCUBA's confusion limit, however they are lower than
previously reported work for $S < 1 \,$mJy.

\begin{table}
\centering
\caption{$N(>S)$ Measurements from this Work.}
\begin{tabular}{lccc}
\hline\noalign{\smallskip}
$S$ (mJy) & $N(>S)$ (deg$^{-2}$) & $\sigma_{N+}$ (deg$^{-2}$) &
$\sigma_{N-}$ (deg$^{-2}$) \smallskip \\
\hline\noalign{\smallskip}
8 & 259 & 41 & 45 \\
6 & 541 & 68 & 74 \\
4 & 974 & 119 & 109 \\
2 & 1950 & 340 & 310 \\
1 & 3730 & 950 & 720 \\
0.5 & 7620 & 2410 & 1810 \\
0.25 & 17300 & 6700 & 5000 \\
\hline
\end{tabular}
\label{tab:meas}
\end{table}

Our catalog surveys a larger set of clusters than any previous work,
with a factor of $\sim 2$ more area at the source plane, and thus
produces statistically tighter error bars. Importantly, the sources in
this catalog are sampled from clusters which were not necessarily
selected for their strong gravitational lensing.  We believe our data
are less biased than some previous lensing surveys.  Several of the
targets -- Abell 2218 and MS$0451.6{-}0305$ are examples -- are known to
contain strongly lensed sources.  Previous work is concentrated on
these and similar fields to yield the largest number of sources per
cluster with a median number of sources per cluster field of $4$.  In
contrast, the catalog used here has a median number of two sources per
cluster field.  Another difference from previous work is the
conservative data analysis and point-source extraction used in
\citet{Zemcov2007} which led to the catalog used here.  In that work,
it was found that many sources from previous analyses drop
significantly below the selection threshold.

To check whether systematic artifacts dominate the difference between
this work and prior number counts, we have divided our catalog in
several ways: into subsamples based on catalogs presented in previous
work (that of \citealt{Smail2002} and \citealt{Cowie2002}), into
targets with integration times longer and shorter than $15 \,$ks; and
the analysis is repeated on these subsets.  We find that the
subsamples are consistent with the results presented for our full
catalog and are lower than those from previous work.  This is
evidence that the subset of very deep or previously studied clusters
have little statistical difference from the overall set.

\begin{figure}
\centering
\epsfig{file=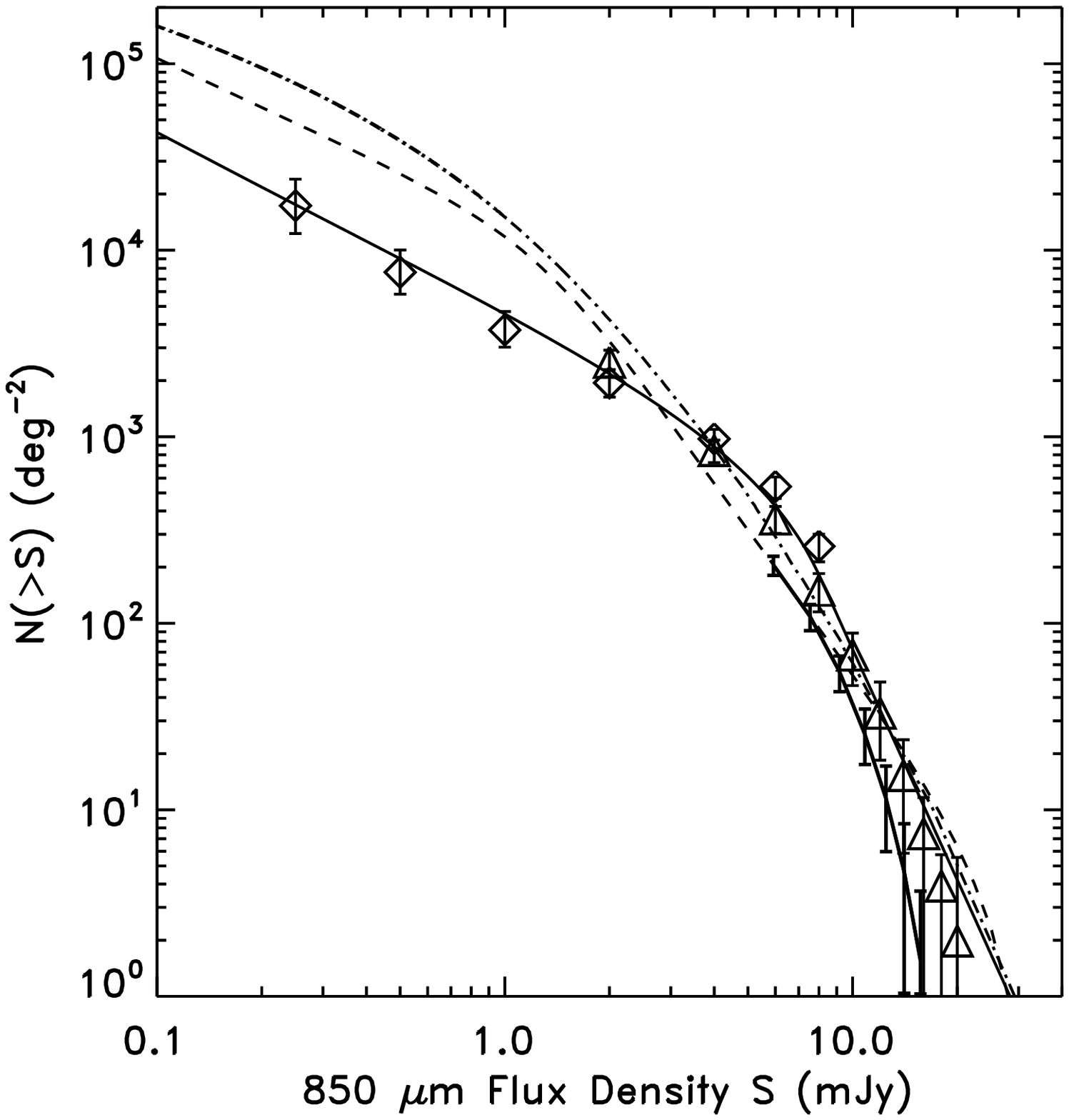,width=0.49\textwidth}
\caption{Comparison of counts measured near $850 \, \mu$m and model
  predictions. The diamonds show the points measured in this work, the
  triangles are the counts measured by SHADES \citep{Coppin2006} and
  the dots are the counts measured by LESS \citep{Weiss2009} where the
  source flux axis has been scaled by a factor $(870 \, \mu \mathrm{m}
  / 850\, \mu \mathrm{m})^{3.5}$, (amounting to a few percent) to
  account for the difference in the centroid wavelengths of SCUBA and
  LABOCA. The solid line shows the best fit of Equation
  (\ref{eq:NgtS}) as described in the text.  The dashed line shows the
  prediction using the model of \citet{Lagache2003}, and the dot-dashed
  line shows the prediction for the model of
  \citet{Valiante2009}. Both models require a break at $\sim 1 \,$mJy
  to obtain a total CIB consistent with the FIRAS result.
  The break in counts at $\approx 8$mJy hinted in the LESS and
  SHADES work and presented here shows that the unverified second break
  in the counts at $1 \,$mJy is not required to match the total CIB
  intensity.}
\label{fig:models}
\end{figure}

The $850 \, \mu$m counts found here are plotted in Figure
\ref{fig:models}\ along with the most precise measurements at brighter
fluxes from SHADES \citep{Coppin2006} using SCUBA and LESS using the
LABOCA instrument at $870 \, \mu$m \citep{Weiss2009}.  Model
predictions from \citet{Lagache2003} and \citet{Valiante2009} are also
shown\footnote{These models are available online at {\tt
    http://www.ias.u-psud.fr/irgalaxies/model.php} and {\tt
    http://www.physics.ubc.ca/~valiante/model}, respectively.}.  The
shapes of the SCUBA and LABOCA points are very similar, though the
LESS source density is a bit lower. The bright end of the lensed
counts presented here are higher than either of the direct surveys,
and we believe that our brightest few points may be biased high due to
misidentification of a handful of cluster member sources as weakly
lensed objects. Our survey is an unbiased count of lensed objects
since clusters are not expected to be particularly aligned with
background sources, but clearly the number density of cluster members
is a biased estimator of the general isotropic source
density. Incorrectly counting cluster members as unit-magnification
background galaxies would inflate the brightest counts above the pure
background-only count locus.
 
The data support a single break in a power-law distribution, as in
Equation (\ref{eq:NgtS}), with the break occurring at $S_{0}= 7$ or 8
mJy.  At the same time, it is clear that the full data sets are not
consistent with each other given the quoted uncertainties.  In fitting
Equation (\ref{eq:NgtS}) to the data, we hold $\beta$ fixed at 5.1,
taken from Coppin et al., choose $S_{0}$ and then fit $N_{0}$ and
$\alpha$ to our five lowest flux data points.  In this procedure, the
choice of $S_{0}$ essentially sets the level of the brightest counts
with $8 \,$mJy running through the SHADES points and $\approx 7 \,$mJy
matching LESS.  Reasonable choices of $S_{0}$ have approximately a 1\%
effect on the inferred total flux of $850 \, \mu$m sources.  Our best
fit for $S_{0}=8 \,$mJy is shown as a solid line in Figure
\ref{fig:models}.
  
\begin{figure}
\centering
\epsfig{file=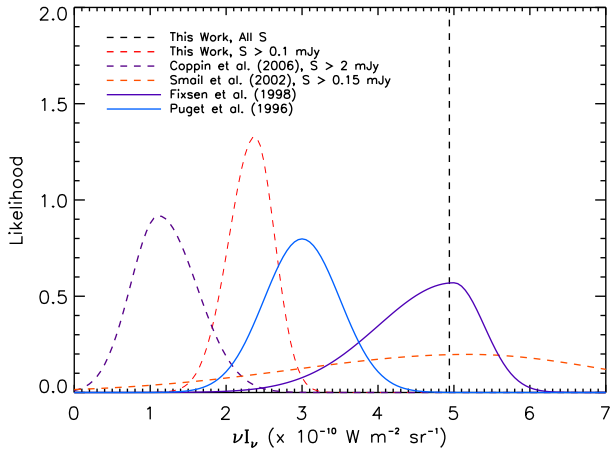,width=0.45\textwidth}
\caption{Likelihoods of selected measurements of the extragalactic
  background at $850 \, \mu$m; those measured by integrating various
  estimates of the FIRAS extragalactic background spectrum over the
  SCUBA band are shown as solid lines, while lower limits on the CIB
  from integrating the SCUBA source counts are shown as dashed. The
  red dashed line shows the limit on the CIB resulting when the counts
  presented here are integrated to $S \geq 0.10 \,$mJy; the PDF has a
  width which is the quadrature sum of the uncertainties on the best
  fitting $\alpha$ and $N_{0}$ described in the text.  The dashed
  black line shows the limits which result when the best-fitting
  $N(>S)$ model is extrapolated to $S = 0 \,$mJy. The measurement of
  \citet{Coppin2006}, which integrates the number counts to $S \geq 2
  \,$mJy, and a representative result from earlier SCUBA surveys of
  lensing clusters integrating the counts down to $S \geq 0.15 \,$mJy
  are shown; because their power laws are steeper, these yield much
  larger intensities if their best-fit models are extrapolated to very
  faint flux.}
\label{fig:nuInu}
\end{figure}

The total brightness of the sky contributed by all sources brighter
than $S_{1}$ is
\begin{multline}
I(\nu) = \int_{S_1}^\infty S \frac{dN}{dS} dS = \\
 N_{0} S_{0}^{2} \left[ \frac{1}{\beta - 2} + \frac{1 - (S_{1} / S_{0})^{2 -
 \alpha}}{2 - \alpha} \right]
\label{eq:firb}
\end{multline}
(where $\alpha=2$ would be treated as a special case).  Values of $\nu
I(\nu)$ corresponding to our best-fit value of $\alpha=1.86$ and to
the $1 \sigma$ error limits on $\alpha$ are listed in Table
\ref{tab:modparams} in units of \nW\ for several choices of the lower
limit in flux, $S_{1}$.  A summary is plotted in Figure
\ref{fig:nuInu}.  The {\sc firas} experiment on {\sc cobe}
\citep{Mather1994} measured the total extragalactic background in the
SCUBA band to be in the range $3 - 5 \times 10^{-10} \,$W m$^{-2}$
sr$^{-1}$ (\citealt{Fixsen1998}; \citealt{Puget1996}).  Given that
only approximately half the total intensity of the CIB at $850 \,
\mu$m comes from sources with $S\geq 100 \, \mu$Jy accessible to SCUBA
lensing surveys, our constraints on the total brightness of the CIB
from sources do not add to the precision available from FIRAS for
the total intensity of the CIB.  At the same time, it is clear that
the data do not require any break in the observed low-flux power law
in order to be consistent with the FIRAS limits.  Previous source
number count determinations using SCUBA have the property that the
extrapolation of $N(>S)$ over all fluxes overproduces this total CIB
by a significant amount; in order to avoid this problem, either a
second sharp break or a shallower faint count index would have been
needed.

\begin{table*}[htdp]
\caption{CIB Fluxes Varying $\alpha$.}
\begin{center}
\begin{tabular}{cccccc}
\hline
$\alpha$ & $N_o$ & $\nu I$ all $S$ & $S \geq 10 \mu$Jy &  $S \geq
100 \mu$Jy &  $S \geq 250 \mu$Jy \cr  
& (deg$^{-2}$ mJy$^{-1}$) & (\nW) & (\nW) & (\nW) & (\nW) \cr 
\hline
1.78 & 98.0 & 0.340 & 0.269 & 0.221 & 0.194 \cr
1.86 & 92.0 & 0.494 & 0.309 & 0.238 & 0.203 \cr
1.92 & 86.8 & 0.848 & 0.364 & 0.266 & 0.222\cr
\hline
\end{tabular}
\end{center}
\label{tab:modparams}
\end{table*}

Predictions given by the models of the infrared background source
population shown in Figure \ref{fig:models}\ qualitatively reproduce
the observed source density, steep slope, and break in the power
law. However, they exhibit a break at $\sim 1 \,$mJy which follows
from matching the evolution of the sources to the counts throughout
the infrared regime. Altering simply the luminosity evolution terms in
these models could match the faint counts but would not move the break
in the power law to match the data. More thorough exploration of the
combined effects of the luminosity and density evolution to understand
disagreement between models and observations should be addressed in
future work.

In addition to the counts determined here at $850 \, \mu$m, source
counts from the BLAST instrument working at shorter sub-mm wavelengths
also show a flattening of the faint end of the curve
\citep{Patanchon2009}.  Furthermore, deep observations of a lensing
cluster with LABOCA independently suggest that the $S_{870 \, \mu
  \mathrm{m}} < 2 \,$mJy part of the $N(>S)$ curve has a flatter slope
than previously measured \citep{Johansson2010}.  These results
corroborate our suggestion that the slope of the faint end of the
sub-mm source counts has been overestimated.

The key to future measurements attempting to resolve the $850 \, \mu$m
background into individual sources will be to survey large numbers of
galaxy clusters to find rare, highly magnified images of very dim
sources.  Experiments such as SCUBA-2 \citep{Holland2006} will allow
mapping of $>100$ cluster fields to the confusion limit of the JCMT in
a matter of weeks of observing time.  This can be compared to the
performance expected from ALMA, which will have much better angular
resolution than any single-dish instrument but a very small field of
view.  At $350 \,$GHz, ALMA can generate a mosaicked map of the
central $r=2'$ of a cluster in about 100 pointings, each requiring
$\sim 3$ minutes to complete.  In contrast, SCUBA-2 can make confusion
noise dominated maps of clusters far beyond this central region with
an instrument noise of $0.5 \,$mJy in $\sim 1$ hr.  Per cluster, this
is a raw mapping speed advantage of a factor of 5 over ALMA on the
central area; a survey strategy which would be maximally efficient
would be to find bright sources with SCUBA-2 and follow up the central
regions near the caustic lines with ALMA when interesting sources have
been identified.

Should a survey from a bolometric camera find bright sources in many
clusters, it will be difficult to apply very detailed lens models to
all of them, particularly ones found in upcoming millimeter SZ effect
surveys which may not have ancillary data on which to rely.
Statistical approaches like or more mature than the one adopted in
this work will be useful for analyzing those data.  More
effort to create high fidelity lensing models -- even if only accurate
on average -- based on simple observable quantities is certainly
justified.  

That said, precise mass models of individual clusters which reduce the
uncertainties on the source plane properties of highly magnified
sources remain a very important component of these studies.  Simple
models are not very accurate when studying individual sources, so it
is certainly worth expending the time and effort required to generate
detailed lensing models.  Combined with such models, the data produced
by very high angular resolution instruments like ALMA will surely
allow us the best view of the faintest galaxies making up the sub-mm
background.


\section*{Acknowledgments}
\label{ack}

M.Z.'s research was supported in part by a NASA Postdoctoral Fellowship.
Many thanks to K.~Coppin for the kind sharing of many of the data used
in Figures \ref{fig:NgtS}\ and \ref{fig:models}, E.~Jullo for help
with {\sc lenstool} and other useful discussions, E.~Valiante for
valuable insights which improved this paper, and A.~Conley for
catching problems quickly.  Thanks also to an anonymous referee whose
comments and suggestions substantially improved this work.  This work
made use of the Canadian Astronomy Data Centre, which is operated by
the Herzberg Institute of Astrophysics, National Research Council of
Canada.


\nocite{Struble1991,Natarajan2004,Fischer1997}
\bibliography{ms}


\end{document}